\documentclass[conference]{IEEEtran}
\usepackage{amsmath,amsfonts}
\usepackage{algorithmic}
\usepackage{graphicx}
\usepackage{textcomp}
\usepackage{xcolor}
\usepackage{booktabs}  
\usepackage{multirow}
\usepackage{colortbl}
\usepackage{comment}
\usepackage{xspace}
\usepackage{pifont}
\usepackage{adjustbox}
\usepackage{makecell}
\usepackage{hyperref}
\usepackage[most]{tcolorbox}

\providecommand{\Description}[1]{}

\colorlet{lightgray}{gray!30}

\newcommand{\README}{README~}
\newcommand{\rqone}{RQ$_1$: How do single-agent and multi-agent architectures compare with LARCH in terms of \README quality and computational cost?}
\newcommand{\rqtwo}{RQ$_2$: How do the generated \README files compare with the original ones in terms of taxonomic coverage?}

\AtBeginDocument{%
	}

\begin{document} 

%
%
%
%
%
%
%
%


\title{The Illusion of Agentic Complexity in README.md Generation: Evaluating Single-Agent vs. Multi-Agent RAG Systems}



\author{%
\IEEEauthorblockN{%
Abu Saleh\IEEEauthorrefmark{1}\IEEEauthorrefmark{2},
Tesfay Welegebreal Tesfay\IEEEauthorrefmark{1}\IEEEauthorrefmark{2},
Phuong T. Nguyen\IEEEauthorrefmark{1},\\
Juri Di Rocco\IEEEauthorrefmark{1},
Muhammad Umar Zeshan\IEEEauthorrefmark{1}\IEEEauthorrefmark{3},
Davide Di Ruscio\IEEEauthorrefmark{1}
}

\IEEEauthorblockA{%
\IEEEauthorrefmark{1}University of L'Aquila, L'Aquila, Italy\\
\IEEEauthorrefmark{2}{\AA}bo Akademi University, Turku, Finland\\
\IEEEauthorrefmark{3}University of Pisa, Pisa, Italy\\
\{abu.saleh, tesfaywelegebreal.tesfay, muhammadumar.zeshan\}@student.univaq.it,\\
\{phuong.nguyen, juri.dirocco, davide.diruscio\}@univaq.it}}

\maketitle

\begin{abstract}
Large Language Models (LLMs) are increasingly utilized to automate several software engineering tasks, including code completion, code summarization, testing, and the generation of repository-level documentation. While Multi-Agent Systems (MAS) are often adopted to support such tasks under the premise that task decomposition improves performance, the impact of architectural complexity on practical efficiency remains under-examined. This study empirically evaluates Retrieval-Augmented Generation (RAG) dependent architectures for the generation of \README files for GitHub repositories. In this work, we conducted a systematic comparison between a Single-Agent pipeline, a specialized MAS, and a developer-guided planning (Dev-Plan) variant, benchmarked against LARCH--a state-of-the-art baseline--and the original ground truth. Results indicate a critical architectural trade-off: the Single-Agent pipeline achieves lexical quality comparable to MAS while reducing token consumption by 86\% and operating at twice the speed. In contrast, manual taxonomy analysis demonstrates that MAS achieves high structural consistency (98\%), resolving formatting issues observed in single-agent approaches. Autonomous planning is identified as the primary pipeline bottleneck; incorporating lightweight developer-guided plans produces the highest overall documentation quality, surpassing all the analyzed configurations.
\end{abstract}

\section{Introduction}
\label{intro}

Software repositories hosted on platforms such as GitHub rely heavily on \README files to communicate project purposes, installation procedures, usage examples, and contribution guidelines. As the primary entry point for users and contributors, \README plays a critical role in project discovery and adoption. However, \README writing remains a manual and often inconsistent process, frequently resulting in incomplete, outdated, or poorly structured documentation \cite{gao2026doesreadmefileneed}.

Recent advances in Large Language Models (LLMs) have significantly improved the ability to generate natural-language text from source code, enabling automated documentation at scale \cite{achiam2023gpt}. Empirical studies show that LLMs can generate code documentation with improved readability and semantic accuracy compared to traditional approaches \cite{dvivedi2024comparative, sarker2025automated}. These capabilities have motivated growing interest in applying LLMs to higher-level documentation tasks, including repository-level summarization and generation of \README files.

Koreeda et al. \cite{koreeda2023larch} proposed LARCH, a heuristic-guided pipeline for automatic \README generation, whereas RepoAgent by Luo et al. \cite{luo2024repoagent} introduces a framework that leverages repository structure and inter-file dependencies to improve documentation coverage. More recent approaches, such as RepoSummary \cite{zhu2025reposummary}, emphasize feature-oriented summarization to better align generated documentation with user-relevant functionality. Additionally, frameworks like ReadMe.LLM \cite{wijaya2025readme} suggest that documentation itself can be optimized to better support both human users and machine understanding.

There has been an increasing interest in multi-agent LLM systems. Frameworks such as AutoGen \cite{wu2024autogen}, MetaGPT \cite{hong2023metagpt}, or ChatDev \cite{qian2024chatdev} decompose software engineering (SE) workflows into interacting roles, enabling iterative refinement and division of labor. This paradigm assumes that tasks such as \README generation benefit from modular task decomposition into subtasks. However, recent evidence suggests a single capable LLM can achieve high-quality repository-level reasoning simply by incorporating structural and dependency cues into a unified context window~\cite{shrivastava2023repository}.

Despite the contrasting nature with their own advantages and limitations, single-agent and multi-agent paradigms have not been empirically compared on the specific task of \README generation. Such an investigation is particularly important because \README files require not only accurate summarization but also coherent organization, consistent style, and user-oriented presentation properties that may be sensitive to fragmentation across multiple agents \cite{gao2025single, nguyen2025teamwork}.

This work supports a call to critically evaluate the necessity of agentic complexity in SE 
with the following contributions:

\noindent $\rhd$ \textbf{Comparative Framework.} A controlled experimental setting was designed to investigate single- and multi-agent approaches for \README generation. Both approaches were evaluated under identical repository contexts to determine whether modular decomposition outperforms monolithic LLMs. To the best of our knowledge, this work represents the first systematic comparison of agentic architectures for automated \README generation.

\noindent $\rhd$ \textbf{Evaluation.} An evaluation of the generated documentation was conducted using standardized assessment methodologies, including manual human inspection. This evaluation explicitly focuses on macro-level coherence, structural consistency, and overall quality of repository-level documentation. 

\noindent $\rhd$ \textbf{Comparison.} We benchmarked the single-agent and multi-agent approaches against LARCH~\cite{koreeda2023larch}, the current state-of-the-art technique, and against the ground truth represented by the original \README files. 

\noindent $\rhd$ \textbf{Open Science.} A complete replication package with code and data has been published to facilitate future research~\cite{ml4se_replication_package}.

    
    
    
    


\section{Related Work}
\label{background}

Automatic \README generation has gained attention due to its role in improving software usability and reuse. More recent work has explored automatic \README generation using LLMs \cite{cui2025rmgenie, koreeda2023larch}. LARCH \cite{koreeda2023larch} is an LLM-based system that generates \README files by selecting representative code fragments using heuristic methods and synthesizing them into natural language descriptions. LARCH demonstrates that grounding generation on salient code elements significantly improves \README quality, but relies on fixed heuristics and single-pass generation. RMGenie \cite{cui2025rmgenie} advances this direction by introducing an agent-based framework for \README generation, where an LLM agent iteratively analyzes repository structure, invokes tools, and refines \README content through multi-step reasoning. 
In addition to agent- and LLM‑based approaches, fine‑tuning and retrieval‑augmented strategies have been applied to automated \README generation. For instance, ReadmeReady \cite{chakrabarty2025readmeready} uses fine‑tuned open‑source LLMs and code indexing to produce basic \README content. 

Multi-agent approaches have also been employed for software documentation tasks. DocAgent \cite{yang2025docagent} leverages multiple specialized agents to collaboratively generate code documentation, demonstrating improved completeness and factual accuracy through iterative reasoning and verification. 
Metagente \cite{nguyen2025teamwork} deploys LLM-based MAS to 
GitHub \README summarization, where cooperating agents iteratively evaluate and refine summaries, outperforming single-agent baselines. 

The usage of LLMs has been also investigated for code summarization and automated documentation \cite{oskooei2025repository, jiang2026survey}. Prior work has shown that LLMs can effectively support code understanding by generating natural language explanations and answering questions about code behavior \cite{leinonen2023comparing, nam2024using}. Early work such as \cite{khan2022automatic} demonstrated that GPT-3 can generate function- and module-level documentation across multiple programming languages. Other approaches, such as context-aware summarization, use broader program structure to generate more accurate and informative descriptions \cite{su2024context}. At the repository level, RepoAgent \cite{luo2024repoagent} introduces an LLM-based framework for generating and maintaining structured code documentation through global repository analysis. 

Unlike prior work, which focuses on code documentation \cite{yang2025docagent} or summarization \cite{nguyen2025teamwork}, our approach aims to compare single- and multi-agent–based approaches to generate \README files starting from the source code of the projects.

\section{Methodology}
\label{sec:methodology}

\begin{figure}[t]
  \centering
  \includegraphics[width=0.98\columnwidth]{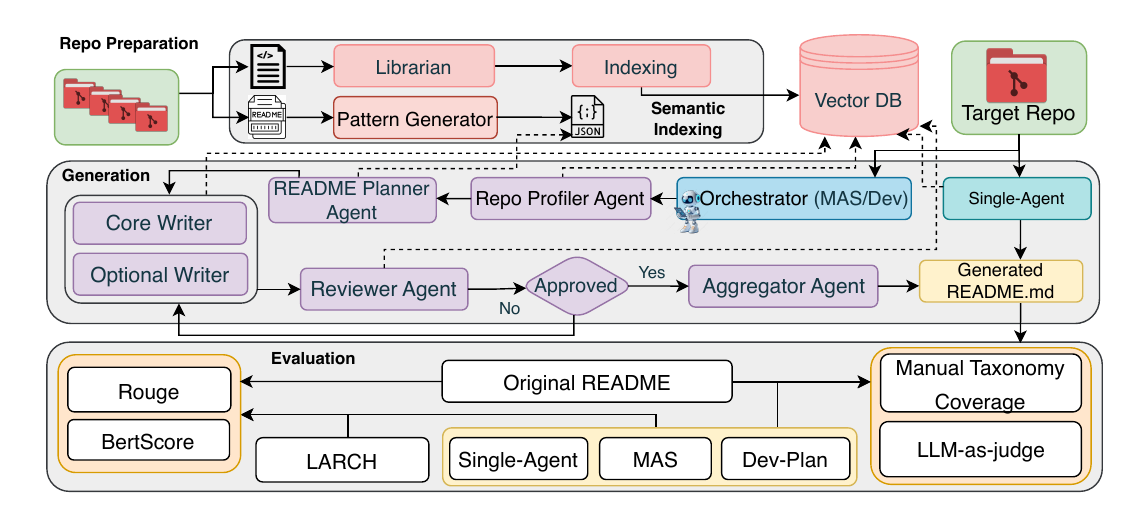} 
  \caption{Overview of the proposed \README generation pipeline, from repository preparation and indexing to generation and evaluation. 
  }
  \label{ML4SE_System}
\end{figure}

In this section, we describe the process that we developed to support the comparison of single and multi-agent systems for the generation of README files. Figure~\ref{ML4SE_System} presents an overview of the process, which includes the \texttt{Repository Preparation}, \texttt{Semantic Indexing}, \texttt{Generation}, and \texttt{Evaluation Pipelines},  described below.\footnote{Due to the strict space limit, we cannot show the prompts and representative intermediate artifacts here. Interested readers are kindly referred to the replication package available online \cite{ml4se_replication_package}.} 



\subsection{Repository Preparation}
\label{repo_preparation}
Before generation, the framework prepares the repository for downstream processing and extracts structural patterns from original \README files.

\noindent $\rhd$ \textit{Acquisition and Isolation}. The original \texttt{README.md} is extracted as ground truth and excluded from the pipeline to prevent access to it during generation. All other Markdown files are similarly removed prior to vector database construction. This prevents system from accessing reference documentation, ensuring that generated READMEs are derived solely from source code and configuration files.

\noindent $\rhd$ \textit{Pattern Extraction}. A canonical \README structure is derived by extracting \textit{H1}, \textit{H2} headings from ground-truth \README. These headings are aggregated across repositories and analyzed using an LLM to identify frequency and semantic patterns. The final output is a JSON file 
that presents a taxonomy of section headers and their recommended sequence.

\subsection{Semantic Indexing}
To distill raw source code into a queryable format without overwhelming LLM context windows, we employ an intelligent, programming language-aware ingestion pipeline.

\noindent $\rhd$ \textit{Intelligent File Selection}. 
A static scanner generates a repository file tree, which is evaluated by a specialized \texttt{Librarian} agent, which isolates essential files (e.g., entry points, configuration manifests, core logic), filtering out noise to significantly reduce embedding overhead.

\noindent $\rhd$ \textit{Segmentation and Indexing}. 
To preserve structural integrity during ingestion, code files are dynamically mapped to their programming language and partitioned using syntax-aware recursive splitters. The splitter divides chunks at logical semantic boundaries (e.g., classes or function blocks) rather than arbitrarily mid-statement. Generic text and configuration files default to a standard character splitter. These chunks are vectorized and persisted in isolated collections within a local vector database. Following prior RAG methods for code comprehension~\cite{cui2025rmgenie, luo2024repoagent}, downstream agents query this index using diversity-aware semantic search, ensuring a comprehensive, non-redundant representation of the codebase.

\subsection{Generation Pipelines}
Utilizing the previously constructed vector databases to provide repository-level context, we implemented single-agent and multi-agent generative pipelines. Furthermore, a human-in-the-loop phase is introduced within the multi-agent systems to evaluate the impact of human oversight on output quality.

\subsubsection{Single-Agent Pipeline} 
To test whether monolithic generation can succeed given sufficient context, the first paradigm implements a unified RAG approach. The system performs a predetermined set of semantic retrieval queries on the vector store. These queries are based on the section taxonomy generated during the pattern extraction phase, with each query formulated as a keyword-level expansion of its respective section type. The retrieved code chunks are concatenated into a single context window. A single LLM is then prompted to generate the fully formatted \README  with one single query.

\subsubsection{MAS Pipeline}
In addition, we designed a multi-agent workflow supervised by a central \texttt{Orchestrator} that routes execution through four phases:


\noindent $\rhd$ \textit{Profiling and Planning}. 
The \texttt{Repo Profiler} agent generates a structured project profile by combining the repository file tree with retrieved context from its repository-specific vector database. The profile summarizes README-relevant information such as project type, dependencies, commands, and key features. Afterwards, the \texttt{\README Planner} agent combines this profile with the \textit{\README Pattern} extracted during the pattern extraction step to produce a JSON file specifying the ordered section structure of the \README and section-level instructions, which is used to guide writer agents.


\noindent $\rhd$ \textit{Section-Level Generation}. The \texttt{Orchestrator} delegates drafting tasks to specialized writing agents: a \texttt{Core Writer} for technical sections (e.g., usage, installation) and an \texttt{Optional Writer} for administrative sections (e.g., licenses). Each writer performs localized retrieval using the \texttt{Planner}'s instructions to obtain section-specific source evidence from the repository-specific vector database. 

\noindent $\rhd$ \textit{Iterative Review}. Drafts are routed to a \texttt{Reviewer} agent, which executes queries against the vector database to verify the factual correctness of the generated content. The retrieved source code chunks serve as the ground-truth reference for assessing the technical accuracy of the drafted text. Sections with inaccurate or incomplete content are returned to the writers with a critique for a rewrite.

\noindent $\rhd$ \textit{Artifact Aggregation}. Once all sections pass review, an \texttt{Aggregator} agent merges them into a contiguous document, ensuring stylistic cohesion and stripping out any redundantly generated command blocks.

\subsubsection{Developer-Guided Planning (Dev-Plan)}
We introduced a developer-guided variant of the MAS pipeline to determine whether performance limitations arise from the planning stage, since LLMs frequently struggle with strategic planning in multi-step workflows \cite{wang2024survey}. In this configuration, the \texttt{\README Planner} agent is omitted, and its output is replaced with a human-authored planning artifact. 
The human developer creates a JSON file that adheres to the \texttt{Planner}'s JSON schema, but contains a manually curated sequence of sections designed to meet repository-specific requirements.
This artifact is provided before execution to directly substitutes the \texttt{Planner}'s output. The downstream MAS components (section-level generation, iterative review, and aggregation) remain unchanged. This design enables evaluation of MAS performance under human-planned conditions.

\section{Experiments}
\label{proof_of_concept}

\subsection{Research questions}
\label{research_questions}

We 
answer the following research questions (RQs).

\noindent $\rhd$ \textbf{\rqone~(Efficacy and Efficiency)}: This RQ evaluates the central goal of the paper: comparing a single-agent RAG pipeline, a multi-agent RAG pipeline, and LARCH~\cite{koreeda2023larch} as the state-of-the-art baseline for README generation. Since LARCH generates documentation by summarizing a single heuristically selected code file, we assess whether repository-level RAG pipelines improve effectiveness and efficiency by overcoming the information loss inherent in LARCH's restricted single-file methodology, and whether added multi-agent complexity yields measurable benefits over a simpler single-agent design.

\noindent $\rhd$ \textbf{\rqtwo~(Structural Consistency)}: This RQ evaluates how closely the generated READMEs 
align with the original ones 
in terms of structural completeness and organizational quality. To this end, we compare the outputs of the single-agent and multi-agent approaches against the original README.md files (golden reference) using manual taxonomy-based coverage analysis and LLM-as-a-Judge.

\subsection{Dataset and Evaluation Metrics}
\label{dataset_metrics}


We considered repositories created after August 2025 to reduce data contamination~\cite{caspari2025studying, cheng2025survey}. Following common repositories filtering practices, we retained non-fork, non-archived software repositories with more than 500 stars, disk size between 100 and 10,000 KB, at most 50 source files, and READMEs longer than 1,000 characters; we also excluded tutorials and ``awesome-*'' lists~\cite{borges2016understanding, munaiah2017curating}. The final dataset contains 180 GitHub repositories (118 Python, 49 JavaScript, 13 Go). Though this is a considerably small number, resulting in an evaluation sample size aligned with recent studies~\cite{cui2025rmgenie, yang2025docagent}. From this corpus, we randomly selected 20 repositories for evaluating the developer-guided planning (Dev-Plan) configuration, consistent with qualitative study sizes~\cite{luo2024repoagent, tao2024magis}.

To evaluate generated content, we use ROUGE and BERTScore, which are widely adopted in documentation generation~\cite{hong2023metagpt, qian2024chatdev}. However, these metrics do not capture structural properties such as organizational flow and consistency~\cite{hoang2025codewiki, hu2022correlating}. Addressing this limitation, we define an evaluation taxonomy of 12 commonly recommended README sections~\cite{ml4se_replication_package}, derived from prior literature~\cite{liu2022readme, wang2023study}. This literature-derived taxonomy is used only for evaluation, while the dataset-derived Pattern described in Section~\ref{repo_preparation} is used only for generation. Using this taxonomy as the evaluation criteria, we employ an LLM-as-a-judge~\cite{zheng2023judging}, which better assesses structural quality and usefulness~\cite{yang2025docagent}. READMEs are scored on a 10-point scale based on completeness and adherence to the derived taxonomy. Scores are averaged over four randomized runs~\cite{shi2025judging} to mitigate bias in judging. We additionally perform a manual analysis of coverage of those derived taxonomies, which is reported in the results section.



\subsection{Experimental Settings}

We utilize 
\texttt{gpt-5.1} 
for all generative and reasoning tasks to ensure a fair comparison. 
Both workflow and API interactions are implemented using LangChain \cite{langchain2026}. Source code is partitioned using a language-aware recursive splitter 
to preserve syntactic boundaries. 
Chunks are vectorized via \texttt{text-embedding-3-small} and stored in ChromaDB. 
To guarantee computational tractability in MAS, 
the iterative refinement loop is strictly capped. 
If a drafted section fails the Reviewer agent's critique three consecutive times, the \texttt{Orchestrator} accepts the final attempt. While this guarantees pipeline termination and bounds token costs, we acknowledge it may occasionally retain suboptimal content.

\section{Results and Discussion}
\label{result}

\subsection{Result Analysis}
\label{result_analysis}


\begin{figure*}[htbp]
  \centering
  \begin{minipage}{0.38\textwidth}
    \centering
    \includegraphics[width=\linewidth]{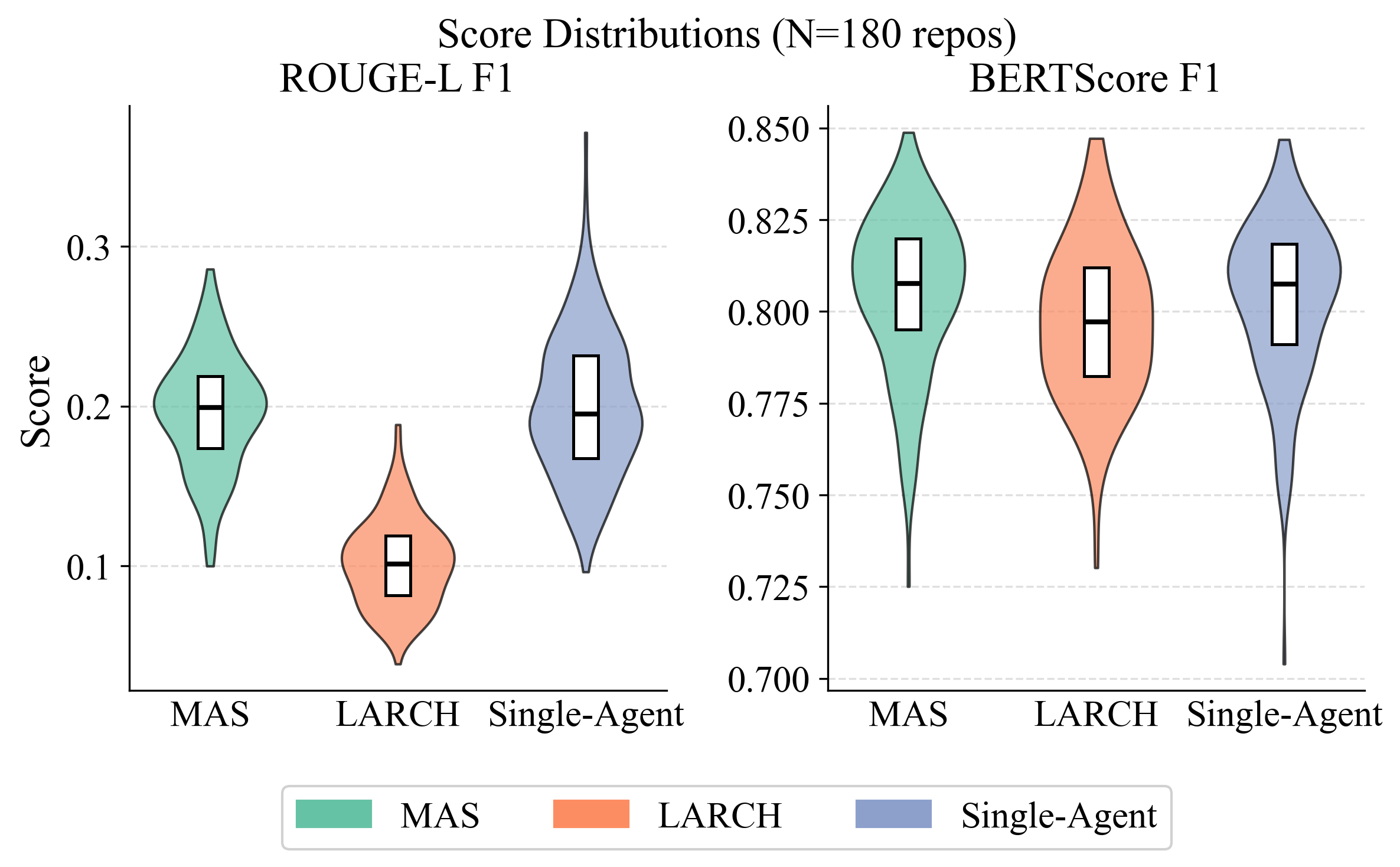}
    \caption{ROUGE/BERTScore: MAS, LARCH, Single-Agent.}
    \label{fig:fig2_violin_eval_3systems}
  \end{minipage}
  \hspace{0.5cm} 
  \begin{minipage}{0.38\textwidth}
    \centering
    \includegraphics[width=\linewidth]{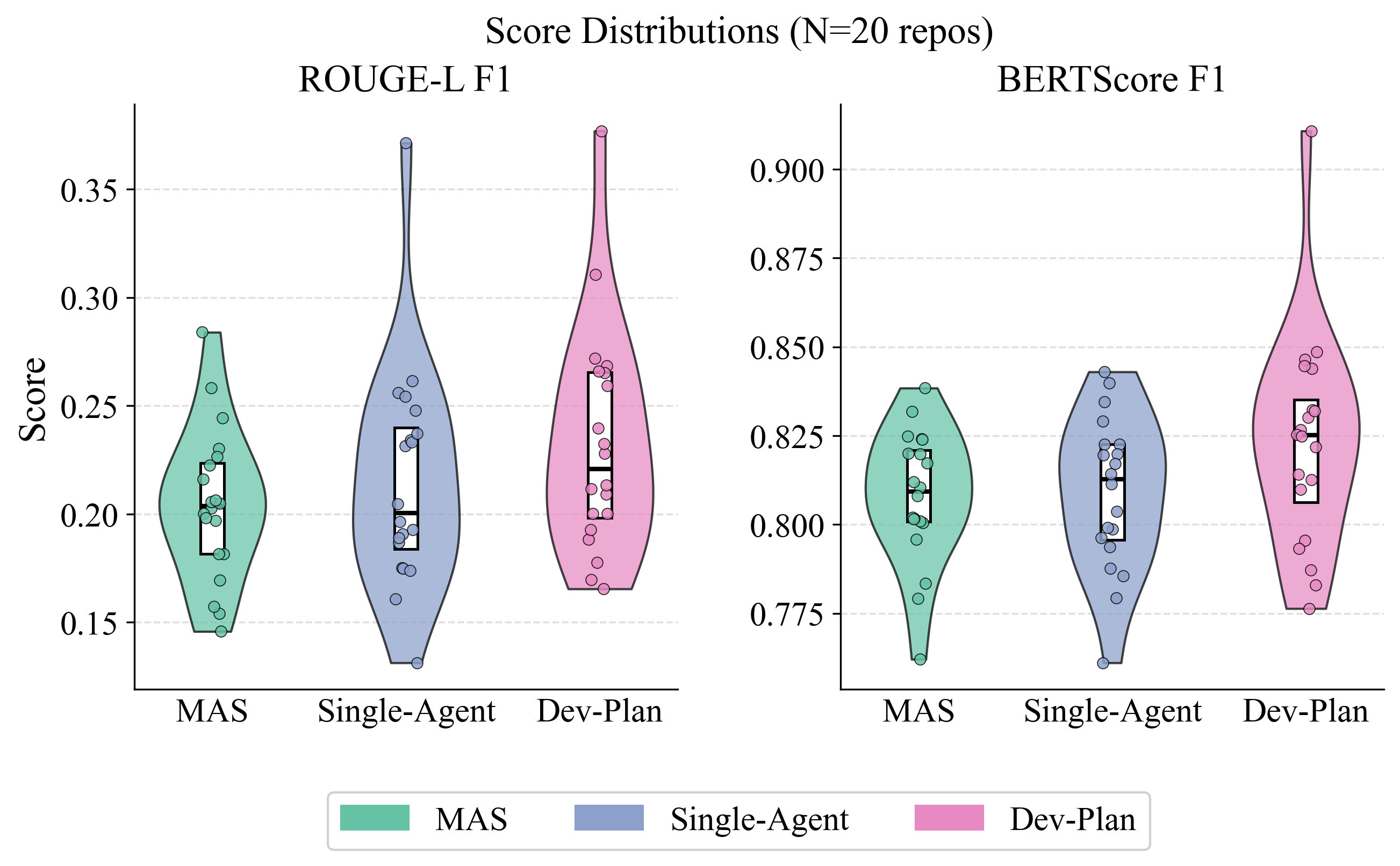}
    \caption{ROUGE/BERTScore: MAS, Dev-Plan, Single-Agent.}
    \label{fig:fig6_violin_eval_3systems}
  \end{minipage}
\end{figure*}

\begin{figure*}[htbp]
  \centering
  \begin{minipage}{0.38\textwidth}
    \centering
    \includegraphics[width=\linewidth]{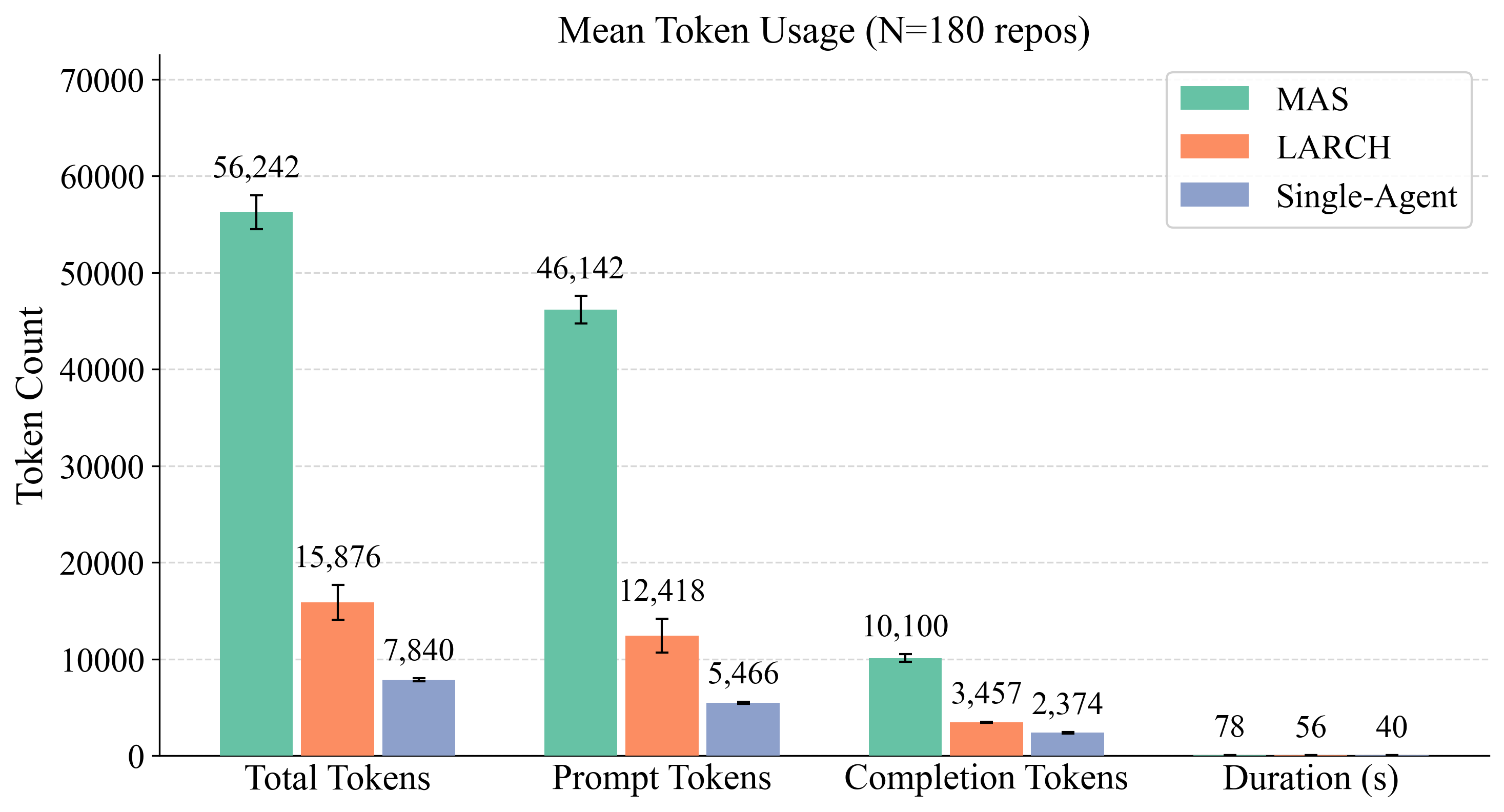}
    \caption{Token usage: MAS, LARCH, Single-Agent.}
    \label{fig:fig3_bar_tokens_3systems}
  \end{minipage}
  \hspace{0.5cm}
  \begin{minipage}{0.38\textwidth}
    \centering
    \includegraphics[width=\linewidth]{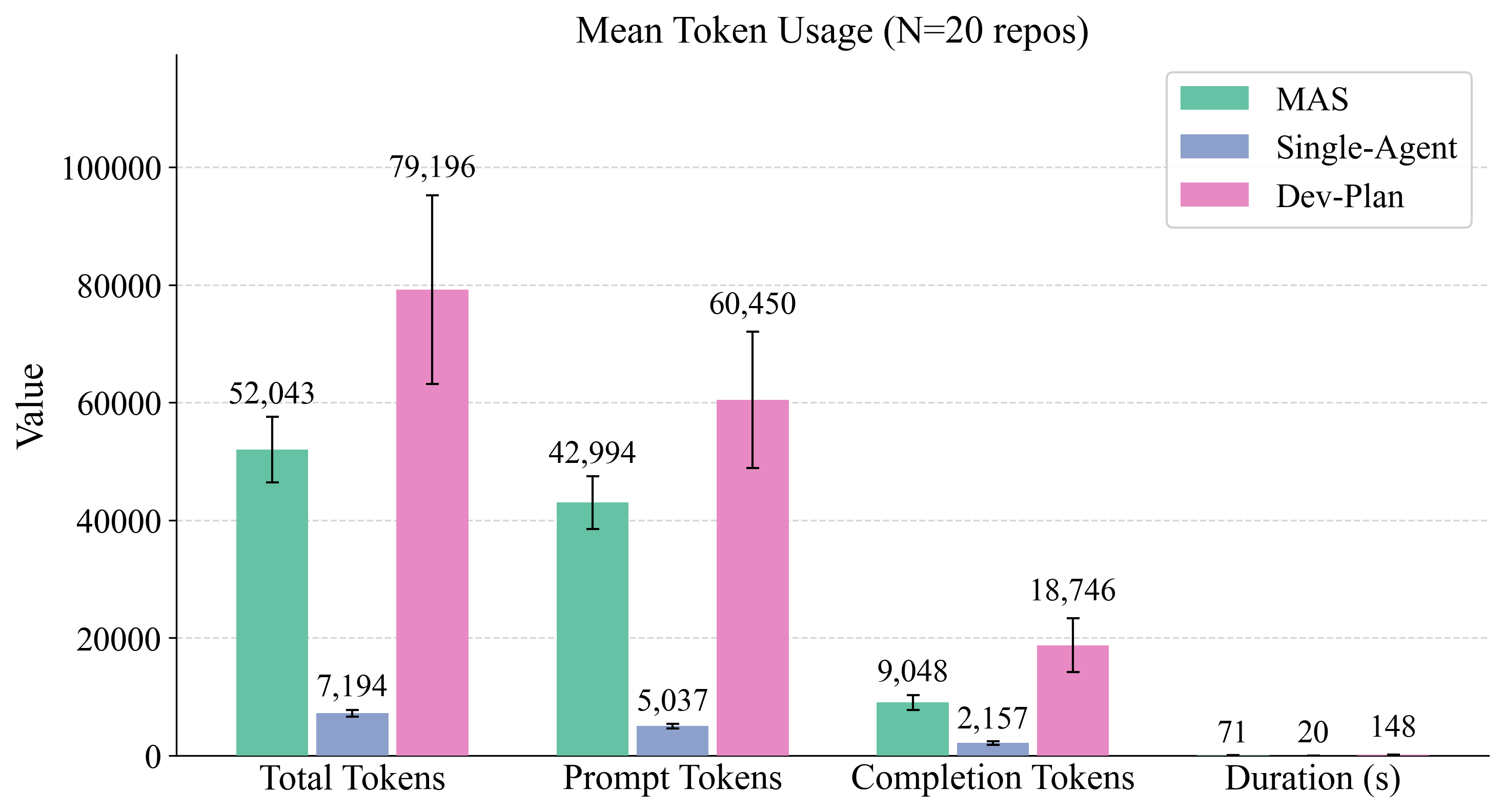}
    \caption{Token usage: MAS, Dev-Plan, Single-Agent.}
    \label{fig:fig7_bar_tokens_3systems}
  \end{minipage}
\end{figure*}

\noindent $\rhd$ \textbf{\rqone}\\
The generated READMEs were evaluated across 180 repositories using ROUGE, BERTScore, and token consumption metrics (Figs.~\ref{fig:fig2_violin_eval_3systems} and \ref{fig:fig3_bar_tokens_3systems}).
The 
experiments challenge the assumption that increased multi-agent complexity inherently produces superior generation. The single-agent pipeline slightly outperforms the fully autonomous MAS in ROUGE-L F1 (0.2007 vs. 0.1964), with negligible differences observed in BERTScore. Both pipelines significantly outperform LARCH. Specifically, LARCH's ROUGE-L F1 decreased substantially to 0.1022. This 
suggests that LARCH's single-file heuristic effectively extracts keywords but does not capture macro-level coherence.

The data indicate a significant operational penalty associated with MAS. Single-agent configuration required an average of 7,840 tokens and 40 seconds to generate each \README file. In contrast, MAS consumed 56,242 tokens, a 7.1x increase driven entirely by prompt overhead, and required 78 seconds. Given the Single-agent's competitive performance, we hypothesized that the autonomous MAS is constrained by cascading errors originating arising from generic AI-generated outlines. To examine this, the Dev-Plan configuration was evaluated, as shown in Fig.~\ref{fig:fig6_violin_eval_3systems}. Dev-Plan achieved the highest overall scores (ROUGE-L F1: 0.2323, BERTScore F1: 0.8230). However, this approach incurs substantial computational cost (Fig.~\ref{fig:fig7_bar_tokens_3systems}). Because human-authored plans contain more granular section specifications, Dev-Plan triggers more section-level retrieval, generation, and review steps, consuming 79,196 tokens and 148 seconds on average.

\begin{tcolorbox}[colback=gray!15, colframe=black, boxsep=0pt, left=3pt, right=3pt, top=3pt, bottom=3pt]
\small{\textbf{Answer to RQ$_1$:} 
Both the single-agent and multi-agent pipelines outperform LARCH in \README generation. The single-agent approach is the most cost-efficient solution, matching MAS quality while using 86\% fewer tokens. However, Dev-Plan successfully resolves the AI-generated outlines limitation in the MAS pipeline and achieves the highest overall quality.}
\end{tcolorbox}

\noindent $\rhd$ \textbf{\rqtwo}

\begin{table}[ht!]
\centering
\caption{Structural Evaluation: Manual Taxonomy Coverage vs. LLM-as-a-Judge ($N=20$).}
\label{tab:structural_eval}
\resizebox{\columnwidth}{!}{%
\begin{tabular}{lcccccc}
\toprule
 & \multicolumn{3}{c}{\textbf{Manual Coverage Analysis}} & \multicolumn{3}{c}{\textbf{LLM-as-a-Judge}} \\ \cmidrule(lr){2-4} \cmidrule(lr){5-7}
\textbf{Method} & \textbf{Precision} & \textbf{Recall} & \textbf{F1-Score} & \textbf{Mean Score} & \textbf{\makecell{1st Place \\ Win Rate}} & \textbf{\makecell{Fail Rate \\(Score $\le 5$)}} \\ \midrule
Dev-Plan & 0.861 & \textbf{0.750} & 0.801 & \textbf{8.60} & \textbf{53.75\%} & \textbf{1.25\%} \\
MAS & \textbf{0.982} & 0.696 & \textbf{0.814} & 7.55 & 16.25\% & 2.50\% \\
Single-Agent & 0.776 & 0.492 & 0.602 & 7.25 & 10.00\% & 6.25\% \\
Golden (Original) & 0.724 & 0.667 & 0.694 & 7.36 & 20.00\% & 17.50\% \\ \bottomrule
\end{tabular}%
}
\end{table}

A dual evaluation was performed on a dataset consisting of 20 repositories (Table \ref{tab:structural_eval}) to assess practical utility beyond n-gram overlap. The initial evaluation consists of a manual analysis of section coverage. For each 
README, we manually mapped its sections to the 12 literature-derived taxonomy, to calculate structural Precision, defined as the proportion of generated sections that match the taxonomy, and Recall, defined as the proportion of taxonomy sections covered. In the second evaluation, an LLM-as-a-Judge assessed overall usefulness on a 10-point scale using the same taxonomy as reference. This combined analysis identifies a significant limitation in standard lexical metrics. While the single-agent approach demonstrated strong performance in ROUGE (RQ1), it underperformed in structural generation, achieving only 49.2\% manual Recall and ranking lowest in LLM-as-a-Judge (Mean: 7.25). Without the strict step by step delegation like multi-agent workflow, single-prompt LLMs produce monolithic, poorly segmented text blocks that omit essential documentation components.

The autonomous MAS demonstrated substantial structural discipline, achieving a manual Precision of 98.2\%. This 
suggests that multi-agent routing effectively eliminates formatting limitations and irrelevant content. The LLM-as-a-Judge metric reflected this organizational strength, assigning MAS higher scores than the Single-Agent and recording a consistently low 2.50\% failure rate. The original README (Golden) was also evaluated against the literature-derived  taxonomy, resulting in lower structural Precision (72.4\%) and a failure rate of 17.50\%, primarily due to project-specific sections and omitted setup or prerequisite instructions. 
Dev-Plan demonstrated the most comprehensive performance. Incorporating human planning enabled the system to achieve the highest structural Recall (75.0\%) and the highest LLM-as-a-Judge scores (Mean: 8.60), while nearly eliminating structural failures (1.25\%).
\begin{tcolorbox}[colback=gray!15, colframe=black, boxsep=0pt, left=3pt, right=3pt, top=3pt, bottom=3pt]
\small{\textbf{Answer to RQ$_2$:} 
Single-agent approaches exhibit limited structural recall, often omitting critical project dimensions. In contrast, the MAS maintains strict adherence to formatting standards, achieving 98.2\% precision. The integration of human strategic planning with AI execution (DevPlan) produces the most comprehensive  documentation. 
These automated systems achieve significantly higher coverage of the elicited taxonomy compared to the \README files authored by human developers.}
\end{tcolorbox}

\subsection{Discussion}
\label{discussion}

\noindent 
\noindent $\rhd$ \textbf{Statistical Validation.} 
Wilcoxon signed-rank test with Holm-Bonferroni correction confirms the superiority of agentic architectures.\footnote{Details of the statistical analysis is available in the replication package \cite{ml4se_replication_package}} Both Dev-Plan and MAS achieved significantly higher F1 scores compared to the Golden human-authored baseline ($p < 0.01$). Furthermore, while Single-Agent performance was comparable to human effort in precision ($p = 0.778$), the MAS approach demonstrated a statistically significant leap in formatting precision ($p < 0.001$), validating the effectiveness of modular decomposition.


\noindent $\rhd$ \textbf{Applicability.} 
The evaluation identifies a critical trade-off: Single-Agent pipelines optimize cost-efficiency, whereas MAS ensures structural completeness, exposing the limitations of lexical metrics for documentation assessment. Notably, the performance of Dev-Plan indicates that AI planning constitutes the primary bottleneck of fully autonomous systems, as generic or inadequate document outlines constrain the quality of downstream drafting agents. 
These findings suggest a necessary shift toward human-AI collaboration. Positioning developers as strategic architects and LLMs as execution engines enables practitioners to balance computational costs while consistently generating industry-standard documentation.

\noindent $\rhd$ \textbf{Limitations.} 
The Librarian agent filters the repository solely based on file names and directory structures to reduce token overhead. This heuristic may exclude essential logic present in files with non-descriptive names. Although the Dev-Plan configuration demonstrates the highest performance, it depends on manually supplied blueprints and was assessed using a smaller dataset. The proposed approach emphasizes structural quality, which may not adequately address the nuances of highly domain-specific documentation. Due to data collection challenges, our evaluation was conducted on 180 repositories. Future work should validate our approach on larger datasets. 

\noindent $\rhd$ \textbf{Threats to Validity.} (i) \textit{Internal validity}: LARCH was adapted to use \texttt{gpt-5.1} instead of its original \texttt{gpt-3} implementation to ensure a fair comparison. (ii) \textit{External validity}: To prevent data leakage, only repositories created after August 2025 were included. Though the findings are valid for this diverse 
dataset, they may not fully generalize to closed-source enterprise monoliths or unconventional codebases.

\section{Conclusion and Future Work}
\label{sec:conclusion}
This study empirically evaluates RAG-augmented architectures for repository-level README generation, questioning the necessity of complex MAS. The findings indicate that a single-agent architecture achieves quality moderately comparable to MAS, reducing token costs, outperforming the baseline. Integrating human planning with AI results in the highest quality. For future work, we plan to study the applicability of agentic architectures in other domains in Software Engineering, such as automated bug fixing or test generation.

	\section*{Acknowledgment} 
	This paper has been partially supported by the MOSAICO project (Management, Orchestration and Supervision of AI-agent COmmunities for reliable AI in software engineering) that has received funding from the European Union under the Horizon Research and Innovation Action (Grant Agreement No. 101189664). This work was partially supported by the Erasmus+ Alliance for Innovation project AIM-PRO--AI Literacy for Multidisciplinary Professional Readiness and Outreach (Grant Agreement No. 101245864) and by Movetia, the Swiss agency promoting exchange, mobility and cooperation in education, continuing education and youth work.

%
%
%
%
%
%


	

\bibliographystyle{plain}
\bibliography{refs}


\end{document}